\documentclass[aps,prb,twocolumn, superscriptaddress]{revtex4}
\usepackage{ulem}
\usepackage{graphicx}
\usepackage{hyperref}
\usepackage{amsmath,amssymb}
\usepackage{bm}
\usepackage{verbatim}

\def\be{\begin{equation}}
\def\ee{\end{equation}}
\def\ba{\begin{array}{lll}}
\def\ea{\end{array}}
\def\ber{\begin{eqnarray}}
\def\eer{\end{eqnarray}}

\newcommand{\dtdrt}{\frac{\text{d}\theta}{\text{d}\tilde\rho}}
\newcommand{\ddtddrt}{\frac{\text{d}^2\theta}{\text{d}\tilde\rho^2}}
\newcommand{\sint}{\sin{\theta}}
\newcommand{\cost}{\cos{\theta}}
\newcommand{\vectomg}{\mathbf{\Omega}}
\newcommand{\vectomgx}{\mathbf{\Omega(\mathbf{x})}}
\newcommand{\DM}{Dzyaloshinskii-Moriya\:}

\begin{document}

\title{Phenomenology of current-skyrmion interactions in thin films with perpendicular magnetic anisotropy}

\author{M.E. Knoester}
\affiliation{Institute for Theoretical Physics, Utrecht
University, Leuvenlaan 4, 3584 CE Utrecht, The Netherlands}

\author{Jairo Sinova}
\affiliation{Department of Physics, Texas A \& M University,
College Station, Texas 77843-4242, USA}

\author{R. A. Duine}
\affiliation{Institute for Theoretical Physics, Utrecht
University, Leuvenlaan 4, 3584 CE Utrecht, The Netherlands}

\date{\today}

\begin{abstract}
We study skyrmions in magnetic thin films with structural inversion asymmetry perpendicular to the film plane. We determine the magnetization texture of a single skyrmion and its dependence on the strength of the \DM interaction relative to the magnetostatic energy. Furthermore, we construct a phenomenological model that describes the interaction between the motion of  
skyrmions  
and electric currents to lowest order in spin-orbit coupling. 
We  estimate the experimental verifiable velocities for current-driven motion of skyrmion textures
based on available  results obtained from domain walls dynamics.
\end{abstract}

\maketitle
\vskip2pc

\section{Introduction}  Topological excitations play an important role in modern physics. \cite{mermin1979} They come in various 
forms, such as cosmic strings, vortices in superfluids and superconductors, and domain walls in ferromagnets. The topological excitations on which we focus here are magnetic skyrmions.

Skyrmion-like configurations of the magnetization direction in ferromagnets 
have been considered in quantum Hall ferromagnets \cite{sondhi1993} and spinor Bose-Einstein condensates. \cite{khawaja2001}  More recently, there has been a surge of interest in ferromagnets with lack of inversion symmetry. The absence of such a symmetry leads, in combination with spin-orbit coupling, to the so-called \DM interactions that favor skyrmion magnetic textures. \cite{bogdanov1994}  This enhanced interest is in large part due to the discovery of skyrmion lattices in MnSi \cite{muehlbauer2009} and other ferromagnets \cite{yu2010} with bulk inversion asymmetry. Moreover, it was subsequently shown that these magnetic textures can be manipulated with charge currents of extremely low densities, \cite{jonietz2010} which leads to attractive possibilities for magnetic-memory applications. \cite{kiselev2011, fert2013}

In these latter experiments, the coupling between skyrmions and electric charge current is largely understood in terms of adiabatic effects in which the spin of the conduction electron adiabatically follows the magnetization texture. Within this picture, the effect of charge current on the magnetization results from the so-called spin-transfer torques exerted by carrier spins on the magnetization. Conversely, the magnetization influences the conduction electrons which leads to effective magnetic fields and the topological Hall effect for static skyrmion textures, \cite{neubauer2009} and effective electric fields for dynamic magnetic textures. \cite{schultz2012}
However, this physical picture neglects intrinsic spin-orbit coupling, in that it assumes that spin-orbit coupling does not affect the dynamical interaction between magnetization and transport current. \cite{hals2013} (Spin-orbit coupling is, however, taken into account via the \DM interactions that determine the equilibrium magnetic texture.)

In the systems on which we focus in this article, the inversion symmetry is broken by interfaces rather than in the bulk and intrinsic spin-orbit coupling is typically important. In particular, we have in mind layered magnetic systems with perpendicular magnetic anisotropy (PMA), \cite{heinze2011,liu2012} such as, for example, Pt/CoFe/MgO and Ta/CoFe/MgO multilayers, that have taken the center stage in experiments on current-driven domain wall motion. \cite{miron2011,haazen2012, emori2013, ryu2013} The motivation for this work is twofold. 

First, recent experiments have shown evidence for \DM interactions in experiments on domain-wall motion in these systems,  \cite{emori2013, ryu2013} opening up the possibility for studying skyrmions as well. Furthermore, we argue that our theory for current-skyrmion coupling is controlled by powers of spin-orbit coupling (via the expansion in magnetization gradients). This enables us to construct a phenomenological model for current-skyrmion interactions that takes into account spin-orbit coupling to lowest order and applies to generic quasi-two dimensional conducting ferromagnets with broken inversion symmetry perpendicular to the plane. This should be contrasted with understanding the coupling between domain walls and current in these systems, which is complicated because of the multitude of torques that can in principle exist, \cite{bijl2012,hals2013b,stier2013} some with different possible microscopic origins. \cite{kurebayashi2013} The theory for coupling of domain walls to current is, however, not straightforwardly controlled by integer powers of spin-orbit coupling. Moreover, detailed microscopic evaluation of the current-induced torques, including all possible effects realistically, is very hard due to the complex nature of the materials involved and the interfaces between them. \cite{haney2013, freimuth2013b} 

The second motivation for studying current-skyrmion coupling is that skyrmions represent a model system for understanding the coupling of current to magnetization.  In addition to current-skyrmion coupling, we study the dependence of skyrmion profile on relative strength of spin-orbit coupling (via the \DM interactions) and magnetostatic energy. Although the systems we have in mind are the PMA materials discussed above, we note that our theory applies to current-driven skyrmion motion in any conducting ferromagnetic system with the above-mentioned inversion asymmetry such as ferromagnets on topological-insulator surfaces, that have attracted attention recently. \cite{garate2010,yokoyama2010, tserkovnyak2012}

Using our theory, we estimate typical skyrmion velocities and find that their order of magnitude is around $10$ m/s for a current density of $10^{11} $ A/m$^2$ (which is a typical current density for experiments on domain wall motion). Below, we first discuss equilibrium skyrmion profiles. Subsequently we study the influence of current on skyrmion motion, and the generation of current by moving skyrmions. We also discuss contributions to the resistivity that arise from current-skyrmion coupling, and, in particular, contributions to the Hall resistivity on top of the contribution due to the topological Hall effect. These extra contributions to the resistivity arise due to the intrinsic spin-orbit coupling, and may be more important than the topological-Hall contribution in the systems that are currently investigated experimentally. 

\begin{figure}
\includegraphics[width=0.5 \textwidth]{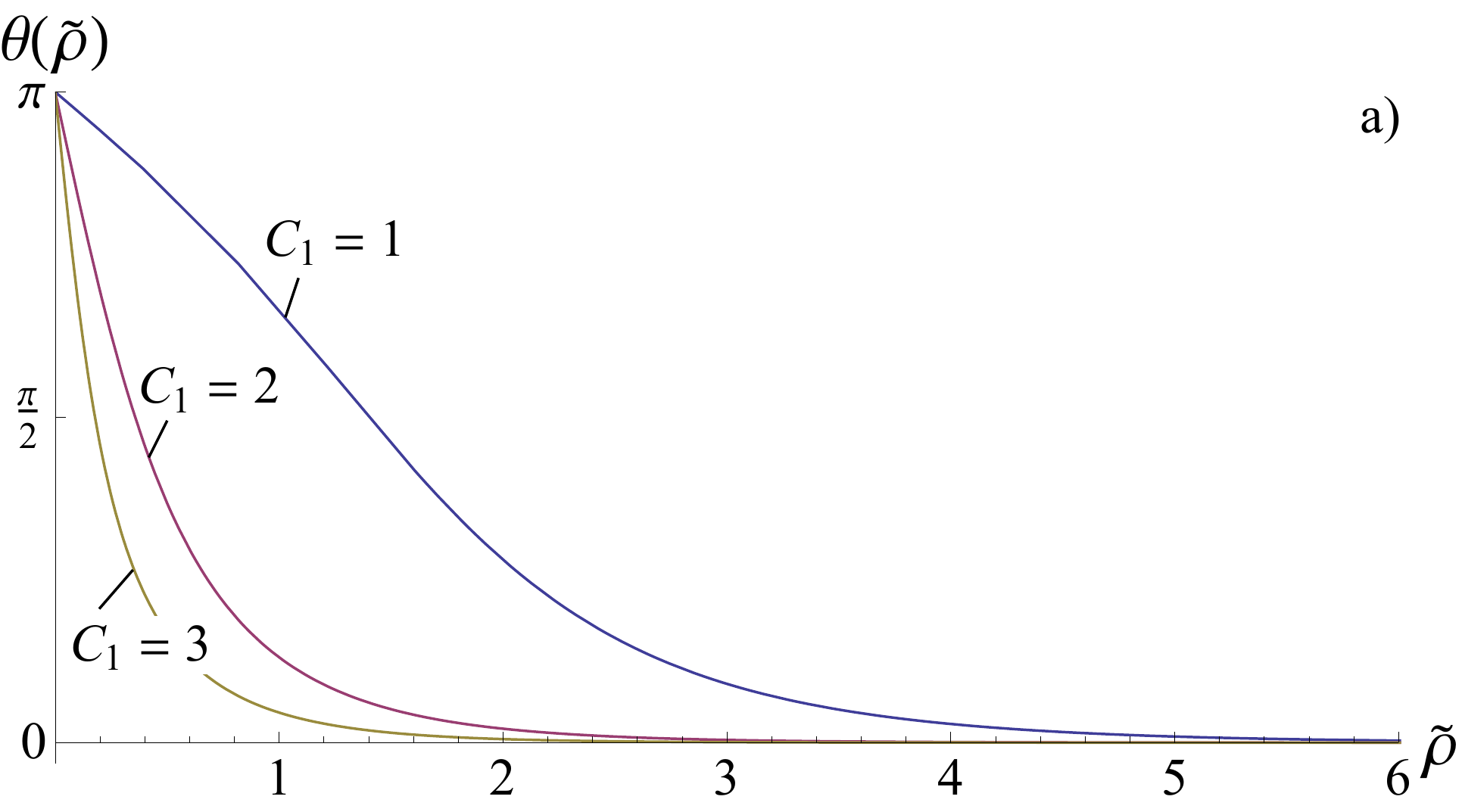}
\includegraphics[width=0.5 \textwidth]{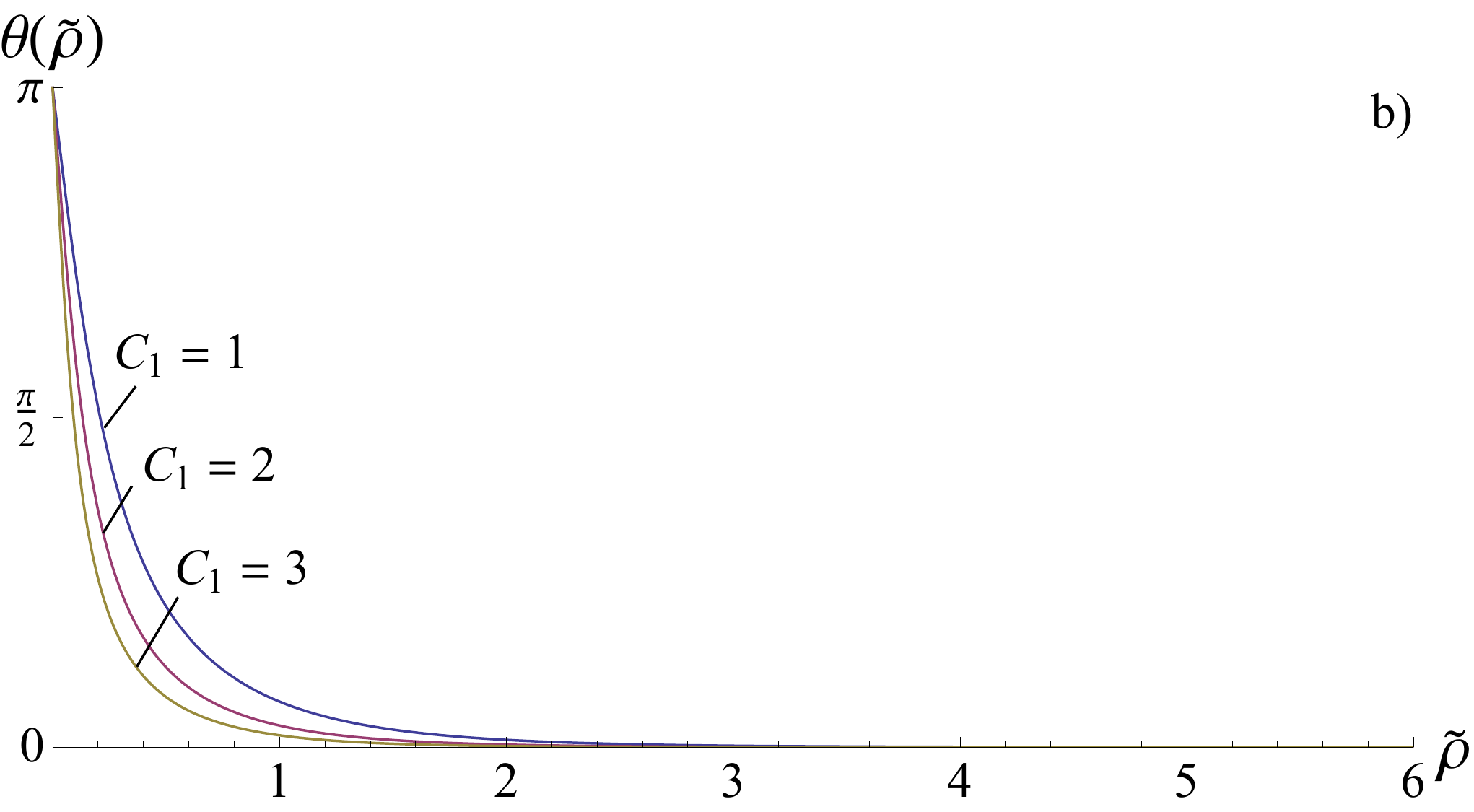}
\includegraphics[width=0.5 \textwidth]{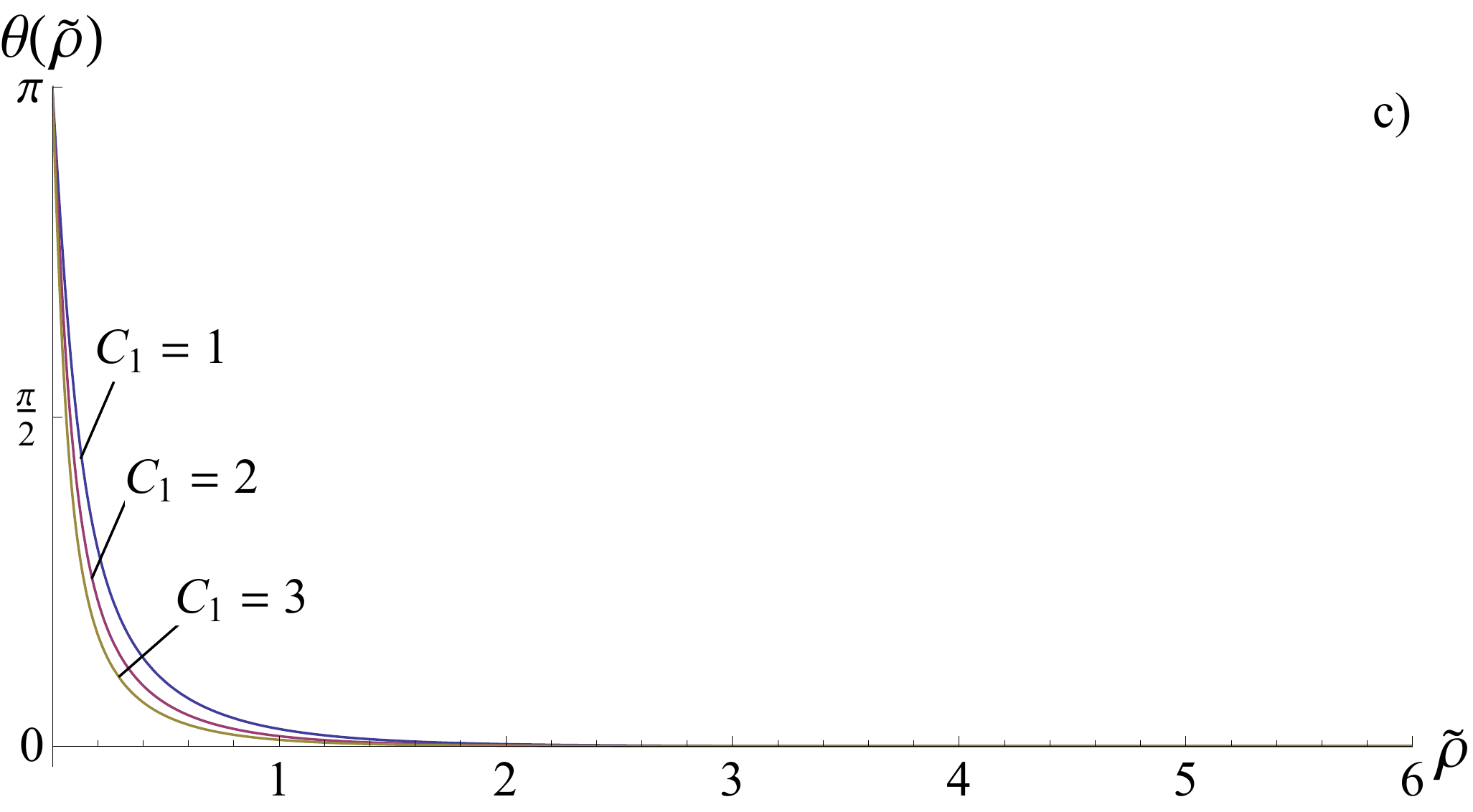}
\caption{Plots of $\theta(\tilde\rho)$ for different values of the parameters $C_1$ and $C_2$. a) $C_2=0$, b) $C_2=1$, c) $C_2=2$. In all cases $C_3=0$.}
\label{fig:profiles}
\end{figure}

\section{Skyrmion profiles} We start out by determining the magnetization texture of a single skyrmion. To second order in the magnetization direction, denoted by the unit vector $\vectomg (\mathbf{x})$  (which is a function of position $\mathbf{x}=(x,y)$ in the plane), and gradients thereof, we have for the energy of the system that
\begin{eqnarray}
\label{eq:energyfirst}
&& E [\vectomg]=t_{\rm FM} \int d\mathbf{x} \left\{-\frac{J_s}{2} \vectomg\cdot\nabla^{2}\vectomg+K(1-\Omega_z^2)\right.\nonumber \\
&& +\left.\frac{C}{2}\left(\hat{y}\cdot\left(\vectomg\times\frac{\partial\vectomg}{\partial x}\right)-\hat{x}\cdot\left(\vectomg\times\frac{\partial\vectomg}{\partial y}\right)\right) \right.\nonumber \\
&&\left.+\mu_0 HM(1-\Omega_z)-\mu_0 M\vectomg\cdot\mathbf{H}_d \rule{0mm}{5mm}\right\}~.
\end{eqnarray}
In the above expression, the thickness of the system in the direction perpendicular to the plane (the $z$-direction) is denoted by $t_{\rm FM}$ and the magnetization direction is assumed not to depend on $z$ (note that $\vectomg$ is, however, a three-dimensional vector). Furthermore, the first term corresponds to the exchange energy with spin stiffness $J_s$ and the second term to anisotropy, proportional to the constant $K$. The \DM interaction is determined by the constant $C$.  That it is indeed related to inversion asymetry in the $z$-direction is made more explicit by noting that the \DM interactions can also be written as $(\hat z \times \vectomg) \cdot (\nabla \times \vectomg)$.  The last two terms in the expression for the energy correspond to external field $H$ (in the $z$-direction) and dipolar field $\mathbf{H}_d$, where $\mu_0$ is the permeability of vacuum and $M$ the saturation magnetization. The dipolar field obeys Maxwell's equations, given by
\begin{subequations}
\begin{align}
\nabla\times\mathbf{H}_d&=0\,~;\label{amaxwell1}\\
\nabla\cdot\mathbf{H}_d&=-M(\nabla\cdot\vectomg)\,~.\label{amaxwell2}
\end{align}
\end{subequations}

To obtain skyrmion profiles we consider solutions with rotational symmetry around the $z$-axis. We write the position vector in cylindrical coordinates, such that $\mathbf{x}=(\rho,\varphi)$, and we consider magnetic textures that are parametrized as follows: $\vectomgx=\sin\theta(\rho)\cos\phi_0\,\hat{\rho}+\sin\theta(\rho)\sin\phi_0\,\hat{\varphi}+\cos\theta(\rho)\,\hat{z}$. Here, the angle $\phi_0$ determines whether the skyrmion is ``hedgehog"-like ($\phi_0=0$), or vortex-like ($\phi_0=\pi/2$). Analogous to Ref.~\onlinecite{bogdanov1994}, but with arbitrary $\phi_0$, we find that the energy of such profiles is \begin{equation}\label{energytilde}
\begin{split}
\frac{E[\theta]}{2\pi t_{\rm FM}}=\frac{J_s}{2}\int&\left\{\left(\dtdrt\right)^2+\frac{\sin{^2\theta}}{\tilde\rho^2}+2C_2(1-\cost)\right.\\
&\left.+\cos{\phi_0}\left(\dtdrt+\frac{\sint\cost}{\tilde\rho}\right)\right.\\
&\left.+(C_1+C_3\cos{^2\phi_0})\sin{^2\theta}\right\}\tilde\rho\,\text{d}\tilde\rho~,
\end{split}
\end{equation}
where $C_1=2 J_s K/C^2$, $C_2=\mu_0 J_s HM/C^2$ and $C_3=2\mu_0 J_sM^2/C^2$ are dimensionless constants that depend on relative strength of \DM interactions and anisotropy, external and dipolar field, respectively, and $\tilde\rho=C\rho/J_s$ is the dimensionless radial position.  In the above, we have taken the system to be translational invariant in the $z$-direction in determining the dipolar field. A detailed treatment of the influence of the finite thickness of the film on the dipolar field and skyrmion size  is beyond the scope of this work. See Ref.~\onlinecite{kiselev2011} for such a study.  Minimizing the energy yields the equation
\begin{equation}\label{difeq}
\begin{split}
&\ddtddrt+\frac{1}{\tilde \rho}\dtdrt-\frac{\sint\cost}{\tilde\rho^2}+\cos{\phi_0}\frac{\sin{^2\theta}}{\tilde\rho}\\
&-(C_1+C_3\cos{^2\phi_0})\sint\cost-C_2\sint=0~.
\end{split}
\end{equation}
The energy contains two contributions related to the angle $\phi_0$. The first is due to the \DM interaction and favors $\phi_0=0$. The second is due to the dipolar field and favors $\phi_0=\pi/2$. We have numerically solved Eq. \eqref{difeq} for various values of $\phi_0$ and $C_3$ and evaluated the energy in Eq. \eqref{energytilde} (which is rather insensitive to the values of $C_1$ and $C_2$) for these solutions. We have found that for values of $C_3$ from zero up to $C_3 \approx 100$, the value of $\phi_0$ that minimizes the energy is $\phi_0 \approx 0$, whereas for larger $C_3$ the angle saturates to $\phi_0 \to \pi/2$. We take the experimental values quoted by Emori {\it et al.} \cite{emori2013}  for which $C_1 \simeq 10$, $C_2 \simeq 1$, $C_3 \simeq 10$. For these parameters we therefore have that $\phi_0=0$. Restricting ourselves to this experimental relevant case we show in Fig.~\ref{fig:profiles} various skyrmion profiles obtained by numerically solving Eq.~(\ref{difeq}). We take the external field to point in the $z$-direction ($H>0, C_2>0$) so that the magnetization at the skyrmion core points in the $-z$-direction, i.e., $\theta (0)=\pi$, whereas the magnetization points in the $+z$-direction sufficiently far from the core. Since the skyrmions are stabilized by  the competition between \DM and exchange interactions, the typical skyrmion size is  on the order of $J_s/C \equiv \lambda$. This is roughly $10$ nm for the experimental values quoted in Ref.~\onlinecite{emori2013}.

\section{Current-driven skyrmion motion} The fact that the scale of the skyrmion is set by the \DM interactions allows for classification of torques that describe coupling between current and skyrmion by their order in spin-orbit interaction, provided this interaction is weak. Introducing $\gamma_{\rm so}$ as the parameter that characterizes the strength of intrinsic spin-orbit coupling, we have that $C \sim \gamma_{\rm so}$, so that for skyrmion profiles each magnetization gradient carries one power of $\gamma_{\rm so}$. In hindsight, this implies that the first and third term in Eq.~(\ref{eq:energyfirst}) are both ${\mathcal O} \left( \gamma_{\rm so}^2 \right)$. Since these determine the skyrmion texture, the energy thus takes into account all possible terms that determine the skyrmion profile to second order in spin-orbit coupling, and including magnetization-direction-dependent exchange interaction is not necessary.

We now proceed by writing down symmetry-allowed current-induced torques and classify them according to their power in spin-orbit coupling. There exist two torques that are also allowed in fully rotation-invariant systems (i.e., system without inversion asymmetry in the $z$-direction), given by
\begin{equation}
\label{eq:sttconventional}
  \left. \frac{\partial \vectomg}{\partial t} \right|_{\rm current} \propto
\left( {\bf j} \cdot \nabla \right) \vectomg + \beta \vectomg \times \left( {\bf j} \cdot \nabla \right) \vectomg~,
\end{equation}
where the current density in the $x-y$-plane is denoted by ${\bf j}$. These are the conventional spin-transfer torques that are commonly used to describe current-driven magnetization dynamics without taking into account intrinsic spin-orbit coupling in current-magnetization interactions. \cite{bazaliy1998,rossier2004,zhang2004} For skyrmions these are ${\mathcal O} \left( \gamma_{\rm so} \right)$ because they are first order in magnetization gradients. The phenomenological parameter $\beta$ determines the relative strength of the reactive and dissipative contribution above that correspond to the first and second term on the right-hand side of Eq.~(\ref{eq:sttconventional}). Note that in microscopic theories  nonzero $\beta$ is the result of extrinsic effects, such as spin-flip scattering, \cite{tserkovnyak2006,kohno2006,piechon2006,duine2007} which we thus treat as being independent of the intrinsic spin-orbit coupling characterized by $\gamma_{\rm so}$. In principle, intrinsic spin-orbit coupling gives a contribution to $\beta$, which would lead to a skyrmion-current coupling term that is second order in spin-orbit coupling and should be neglected in our approach. In the layered systems that we consider here there are, however, always contributions to $\beta$ that do not depend on intrinsic spin-orbit coupling, such as the interface contributions discussed in Ref.~\onlinecite{bijl2013}.  Hence, the term proportional to $\beta$ should be kept within our approximation.

There exist two symmetry-allowed torques that are not expressed in gradients of the magnetization but require intrinsic spin-orbit coupling to occur, \cite{manchon2009,garate2009,pesin2012} and are thus ${\mathcal O} \left( \gamma_{\rm so} \right)$. These are given by
\begin{equation}
\label{eq:shesttandrashba}
  \left. \frac{\partial \vectomg}{\partial t} \right|_{\rm current} \propto
\vectomg \times \left( {\bf j} \times \hat z \right)
+\beta' \vectomg \times \left[ \vectomg \times \left( {\bf j} \times \hat z \right) \right]~.
\end{equation}
One possible interpretation of these torques is that the field-like term (the first term on the right-hand side) is due to a current-induced polarization that exerts a torque on the local moments, with the term proportional to $\beta'$ the associated damping. \cite{kurebayashi2013} Alternatively, the second term on the right-hand side can be interpreted as a Sloncewski-like torque \cite{slonczewski1996} due to absorption (by the ferromagnet) of a spin-Hall-like spin current flowing in the $z$-direction in the normal-metal part of the multilayer with spin polarization in the direction ${\bf j} \times \hat z$, or as an intrinsic anti-damping torque. \cite{bijl2012,kurebayashi2013} The first term is then the associated field-like component. These microscopic interpretations of the torques cannot be distinguished by symmetry and are therefore treated here by a single parameter $\beta'$. Regardless of their microscopic interpretation, both torques in Eq.~(\ref{eq:shesttandrashba}) are  allowed by symmetry and first order in $\gamma_{\rm so}$.

In addition to the four torques in Eqs.~(\ref{eq:sttconventional})~and~(\ref{eq:shesttandrashba}), there exist many more symmetry-allowed torques due to the combined effects  of intrinsic spin-orbit coupling and magnetization gradient that are thus of order ${\mathcal O} \left( \gamma_{\rm so} \nabla \right)$. \cite{bijl2012} For skyrmions, however, these are necessarily  ${\mathcal O} \left( \gamma_{\rm so}^2 \right)$, and can, to lowest order in spin-orbit coupling, be neglected. Hence, to lowest order in spin-orbit coupling the coupling between skyrmion texture and current is described by
\begin{eqnarray}
\label{eq:sttskyrmions}
  && \left. \frac{\partial \vectomg}{\partial t} \right|_{\rm current} =
a\left( {\bf j} \cdot \nabla \right) \vectomg + a' \vectomg \times \left( {\bf j} \cdot \nabla \right) \vectomg \nonumber \\
&&+b \vectomg \times \left( {\bf j} \times \hat z \right)
+b' \vectomg \times \left[ \vectomg \times \left( {\bf j} \times \hat z \right) \right]~,
\end{eqnarray}
where $a, a', b, b'$ are system parameters that can be evaluated microscopically for simple model systems. Given the complexity of the systems under consideration, we treat them here as phenomenological parameters. Note that Eq.~(\ref{eq:sttskyrmions}) includes, to first order in spin-orbit coupling, the current-induced torques both for single skyrmions and skyrmion lattices.

The dynamics of the skyrmions is conveniently studied by means of the Thiele equation, which follows from projecting the Landau-Lifschitz-Gilbert (LLG) equation on the zero mode corresponding to skyrmion motion. \cite{jonietz2010} This approach is valid provided the driving current is small. \cite{makhfudz2012} The LLG equation, including the current-induced torques discussed above, is given by
\begin{equation}
\label{eq:LLG}
 \frac{\partial \vectomg}{\partial t} = -\frac{\gamma}{M} \vectomg \times \frac{\delta E[\vectomg]}{\delta \vectomg} - \alpha_G \vectomg \times \frac{\partial \vectomg}{\partial t} + \left. \frac{\partial \vectomg}{\partial t} \right|_{\rm current}~,
\end{equation}
where $\gamma$ is the gyromagnetic ratio and we have added a Gilbert damping term parameterized by the constant $\alpha_G$. At this point we note that, although we have included all terms that describe coupling between current and skyrmions to first order in intrinsic spin-orbit coupling, the above equation does not contain the anisotropic generalization of gyromagnetic ratio (via the left-hand side of the LLG equation), nor the anisotropic generalization of the Gilbert damping constant. Although these anisotropies are in principle present, they will not affect skyrmion motion at small currents.

As an {\it ansatz} for the LLG equation we take $\vectomg ({\bf x},t) =\vectomg_0 ({\bf x} - {\bf X}(t) )  $, where $\vectomg_0 ({\bf x} )$ is a static skyrmion profile, and ${\bf X} (t)$ the position of the skyrmion. Within this description we then find the Thiele equation (see also Ref.~\onlinecite{sampaio2013})
\begin{eqnarray}
\label{eq:thieleequation}
  &&\epsilon_{\alpha\beta} \left( \dot X_\beta + a j_\beta \right)= \nonumber \\ && - D_{\alpha\beta} \left( \alpha_G \dot X_\beta + a'j_\beta \right) + b \lambda I_{\alpha\beta} j_\beta + b' \lambda I'_{\alpha\beta} j_\beta~,
\end{eqnarray}
where the dot denotes a derivative with respect to time, and $\epsilon_{\alpha\beta}$ is the Levi-Civita symbol, and where summation over repeated indices $\alpha, \beta \in \{x,y\}$ is assumed. In the above, we used that the skyrmion winding number is an integer given by
\begin{equation}
\label{eq:winding}
  W = \int \frac{d{\bf x}}{4\pi} \vectomg ({\bf x}) \cdot \left(  \frac{\partial \vectomg}{\partial  x} \times \frac{\partial \vectomg}{\partial   y}\right)~,
\end{equation}
where in the case of a skyrmion lattice the integration is over one unit cell of the lattice.  The above winding number is associated with the mapping that underlies topological protection of the skyrmion excitation. \cite{mermin1979} In our case we have that $W=-1$. For a single skyrmion we have furthermore that $D_{\alpha\beta} = D \delta_{\alpha\beta}$, $I_{\alpha\beta} = -I \epsilon_{\alpha\gamma}R_{\gamma\beta} (\phi_0)$, and $I'_{\alpha\beta} = - I' R_{\alpha\beta} (\phi_0)$. Here, $D, I, I'$ are dimensionless numbers, \cite{integrals} and $R_{\alpha\beta} (\phi_0)$ are the matrix elements of the matrix performing counterclockwise rotations over an angle $\phi_0$. \cite{matrix} Eq.~(\ref{eq:thieleequation}) also describes rigid translation of a skyrmion lattice. In that case the coordinate ${\bf X}$ is the position of one of the skyrmions and the position of the others follows by lattice translations. Furthermore, the tensors $D_{\alpha\beta}$, $I_{\alpha\beta}$, and $I'_{\alpha\beta}$ are then evaluated by carrying out the appropriate integrals over the unit cell of the skyrmion lattice. \cite{integralslattice}

\section{Charge transport in the presence of skyrmions}  Having discussed the influence of transport currents on skyrmion motion, we turn to the reverse effect, i.e., the current $ {\bf j}^{\vectomg}$ induced by skyrmion motion. Using Onsager reciprocity, \cite{tserkovnyak2008b} we find from Eq.~(\ref{eq:sttskyrmions})~and~(\ref{eq:LLG}) that
\begin{eqnarray}
\label{eq:currentduetotexture}
  && j^{\vectomg}_\alpha = \frac{\sigma M}{\gamma}\left\{ a  \frac{\partial \vectomg}{\partial x_\alpha} \cdot \left( \vectomg \times \frac{\partial \vectomg}{\partial t}\right) - a' \frac{\partial \vectomg}{\partial x_\alpha} \cdot \frac{\partial \vectomg}{\partial t}\right. \nonumber \\ && \left.- b  \left(\hat z \times \frac{\partial \vectomg}{\partial t} \right)_\alpha
+ b'  \left[ \hat z \times \left( \frac{\partial \vectomg}{\partial t}\times \vectomg \right)\right]_\alpha \right\}~,
\end{eqnarray}
where $\sigma$ is the diagonal part of the conductivity to zeroth order in spin-orbit coupling, such that the above equation is second-order in spin-orbit coupling for skyrmions (since $\partial/\partial t \sim \nabla \cdot \dot X \sim \gamma_{\rm so} \dot X$). The first term in the above equation has been dubbed spin motive force, \cite{stern1992,barnes2007} with the second a dissipative correction. \cite{tserkovnyak2008b,duine2008} The last two terms have been derived microscopically  in Ref.~\onlinecite{tatara2013} starting from the Rashba hamiltonian. (See also Ref.~\onlinecite{yamane2013}.)

Skyrmion magnetic textures also give rise to an additional, texture-induced, contribution, to the Hall effect. In Ref.~\onlinecite{neubauer2009}, current-driven skyrmion motion was detected electrically via a drop in this contribution to the Hall effect. This drop is a result of the spin-motive force contribution to the electric field [arising from the first term in Eq.~(\ref{eq:currentduetotexture})], that counteracts the applied electric field. This analysis applies to vanishing intrinsic spin-orbit coupling in the interaction between curent and skyrmions. To investigate how intrinsic spin-orbit coupling alters these texture-induced Hall effects, we consider an applied electric field ${\bf E}$, in addition to the currents induced by skyrmion motion. Hence, we have that ${\bf j} = \sigma {\bf E} + {\bf j}^{\vectomg}$. In the situation of a drifting skyrmion texture (i.e., a single skyrmion or skyrmion lattice) we have that $\partial \vectomg_0 ({\bf x} - {\bf X} (t))/\partial t  = -\dot {\bf X} (t) \cdot \nabla \vectomg$. Assuming that transport is dominated by a single band with carrier density $n$ and carrier charge $e$, we estimate the contributions to the resistivity due to the coupling between current and textures by replacing $\dot {\bf X} \to \dot {\bf X} - {\bf v}$, where ${\bf v} = {\bf j}/ne$
is the carrier drift velocity. Inserting this into Eq.~(\ref{eq:currentduetotexture}) and setting $\dot {\bf X}=0$ allows us to extract the resistivity. We find for the contributions to the resistivity that result from the current-skyrmion coupling in Eq.~(\ref{eq:sttskyrmions}) that
\begin{eqnarray}
\label{eq:resistivity}
  && \Delta \rho_{\alpha\beta} = -\frac{Ma}{\gamma ne} \frac{\partial \vectomg}{\partial x_\alpha} \cdot \left( \vectomg \times \frac{\partial \vectomg}{\partial x_\beta}\right)
 + \frac{M a'}{\gamma ne}  \frac{\partial \vectomg}{\partial x_\alpha} \cdot  \frac{\partial \vectomg}{\partial x_\beta} \nonumber \\
&&
+\frac{M b}{\gamma ne} \left( \hat z \times  \frac{\partial \vectomg}{\partial x_\beta} \right)_\alpha
-\frac{M b'}{\gamma ne} \left[ \hat z \times \left(  \frac{\partial \vectomg}{\partial x_\beta} \times \vectomg \right) \right]_\alpha.
\end{eqnarray}
The first term in the above expression corresponds to the topological Hall resistivity. This Hall contribution is topological as its contribution per skyrmion is determined by the winding number defined in Eq.~(\ref{eq:winding}).   The third term is an extra magnetic-texture-related contribution to the Hall resistivity that arises due to spin-orbit coupling. The second and fourth terms are ordinary, i.e., planar and diagonal, contributions to the resistivity arising from coupling between texture and current. The first of these is present without spin-orbit coupling, \cite{tserkovnyak2009} whereas the other arises because of spin-orbit coupling in combination with magnetization gradients. As a side remark, we note that --- apart from the topological Hall contribution --- the above corrections to the resistivity are nonzero also for textures other than skyrmions, such as domain walls. Also note that there are other contributions to the resistivity resulting from the magnetization, such as the anomalous Hall resistivity, that are not included in $\Delta \rho$. The above contributions to the resistivity result from the current-skyrmion coupling in Eq.~(\ref{eq:sttskyrmions}).

To investigate the current in response to moving skyrmions or a linearly-moving skyrmion lattice, we evaluate  Eq.~(\ref{eq:currentduetotexture}) for a drifting magnetization texture so that
$\partial \vectomg/\partial t  = -(\dot {\bf X} (t) - {\bf v}) \cdot \nabla \vectomg$, where we allowed for the transport current via the drift velocity ${\bf v}$. This yields
\begin{eqnarray}
\label{eq:currentskyrmionansatz}
 \frac{ \gamma A j^\vectomg_\alpha}{4\pi\sigma M} \!= \! \left( a \epsilon_{\alpha\beta} - a' D_{\alpha\beta} +  b \lambda I_{\beta \alpha}  - b' \lambda I'_{\beta\alpha} \right) \left( v_\beta\!-\!\dot X_\beta \right),
\end{eqnarray}
where $A\sim \lambda^2$ is the area occupied by a single skyrmion (in the case of a skyrmion lattice, its unit cell). The terms proportional to $a$ and $I_{\beta\alpha} \sim \epsilon_{\alpha \beta}$ give, for zero skyrmion velocity $\dot {\bf X}=0$ a Hall contribution to the current (the contribution $\sim a$ being the topological Hall contribution, the other being the correction due to spin-orbit couping). Once the skyrmion lattice depins, the Hall current drops because the difference between longitudinal electron drift velocity and skyrmion velocity goes down.  As a final remark, we note that the Joule heating associated with the current excited by the dynamic skyrmion texture [Eq.~(\ref{eq:currentduetotexture})] gives an additional channel for magnetization relaxation that can be quite large in the clean limit, as was pointed out in Ref.~\onlinecite{zang2011}. This latter work did however not take into account intrinsic spin-orbit coupling effects on current-magnetization interactions.

\section{Conclusions and discussion} We have developed a phenomenological model for current-skyrmion interactions to first order in spin-orbit coupling, applicable to metallic ferromagnets with inversion asymmetry in one direction. In addition to adiabatic and non-adiabatic spin-transfer torques, this model takes into account field-like and Slonczewski-like torques that result from the inversion-symmetry breaking in combination with intrinsic spin-orbit coupling. The effects of these torques on skyrmion motion depend on the internal structure of the skyrmions that is determined by the angle $\phi_0$ (this angle is zero for hedgehog-like skyrmions, and $\pi/2$ for vortex-like skyrmions). In particular, the direction of the current-induced forces on the skyrmions, with respect to the current direction, depend on this angle. This dependence could in principle be probed by studying systems with different strengths of \DM interactions.

We point out that the classification of current-induced torques by their order in intrinsic spin-orbit coupling is possible for skyrmions because their size is set by the spin-orbit interaction itself. (In principle, the spin-orbit coupling should be weak; the strength of spin-orbit coupling can, to some extent, be tuned by doping with light elements or varying the composition of the non-magnetic alloy in the multilayer systems.) This implies that all factors involving magnetization gradients give rise to one power of spin-orbit coupling. In addition to being intrinsically interesting, skyrmion textures in PMA materials therefore also provide an important model system for comparing microscopic theories for current-induced torques in these systems with experiment.

To estimate a typical skyrmion velocity, we take parameters from Emori {\it et al.}, \cite{emori2013} for which the dimensionless parameters are $C_1\simeq 10, C_2\simeq 1, C_3\simeq 10$ so that $\phi_0=0$. We then find that $D\simeq 1$, $I\simeq 0.003$, and $I'\simeq -0.007$. Adopting the viewpoint of Emori {\it et al.}, \cite{emori2013} that the Slonczewksi-torque is the dominant one, we have that $a=a'=b=0$, and that $b'=\hbar \gamma \theta_{\rm SH}/(2eMt_{\rm FM})$ with $\theta_{\rm SH}$ the effective spin-Hall angle of the normal-metal layer.  Using the values $t_{\rm FM}=1$ nm, $\theta_{\rm SH}=0.1$ and $M=3 \times 10^5$ A m$^{-1}$, \cite{emori2013} we find that (taking again $\lambda =10$ nm), $|\dot{\bf  X}| \sim 0.1/\alpha_G$ m s$^{-1}$ for current densities of $|{\bf j}| \sim 10^{11}$ A/m${^2}$. Here, we used that in nanostructures the current-induced forces perpendicular to the current are balanced by repulsive forces from the sample edge due to the \DM interactions, leaving the longitudinal forces as the main driving mechanism. \cite{sampaio2013, iwasaki2013, rohart2013} For typical values of the Gilbert damping parameter $\alpha_G \sim 0.1-0.01$ this velocity is of the same order as domain-wall velocities reported for these systems. For the same parameters, we have that voltage drop $\Delta V$ per skyrmion is $\Delta V \sim M b' |\dot {\bf X}-{\bf v}|/\gamma \sim 10$ nV $\times|\dot {\bf X}-{\bf v}|$ [m/s]. Depending on the internal structure of the skyrmion, this voltage drop could be longitudinal to the direction of skyrmion motion ($\phi_0=0$), or transverse ($\phi_0=\pi/2$). As a side remark, we note that for moving domain walls in systems with $b' \neq 0$  a similar voltage drop (longitudinal for N\'e\`el walls, tranverse for Bloch walls) is expected. Hence, this voltage could potentially be used to electrically detect the position and motion of skyrmions and domain walls. Finally, we note that the experimental finding that the adiabatic spin-transfer torque is small ($a \simeq 0$ \cite{emori2013}) implies that the topological Hall effect signal is probably small in the materials with perpendicular magnetic anisotropy that are currently under investigation.

The skyrmions we have considered in this paper could be metastable single skyrmions, nucleated, for example,  by inhomogeneous spin current distributions or by current in the presence of inhomogeneities induced by intended edge defects. \cite{sampaio2013,iwasaki2013,romming2013} Alternatively, they could be thermodynamically stable skyrmion lattices. The latter occur once the energy gain due to \DM interactions is comparable to the energies associated with fields and anisotropy, \cite{bogdanov1994} i.e., when $C^2/J_s \sim K, \mu_0 H M, \mu_0 M^2$. For the parameters reported in Ref.~\onlinecite{emori2013} we have that $C^2/J_sK$ is of order one, so that the prospects of observing skyrmion lattices in these systems appear to be good.

Finally, we mention that intrinsic spin-orbit coupling also results in phase-space Berry phases that affect semi-classical electron dynamics, as discussed for MnSi in Ref.~\onlinecite{freimuth2013}. Since we have worked at a phenomenological level, these effects are not directly visible in our formalism (see, however, Ref.~\onlinecite{bijl2012}). A possible direction for future work would be to study these effects explicitly.

\acknowledgements It is a pleasure to thank Kjetil Hals, Arne Brataas, and Yaroslav Tserkovnyak for discussions. This work was supported by the Stichting voor Fundamenteel Onderzoek der Materie
(FOM), the Netherlands Organization for Scientifc Research (NWO),  by the European Research Council
(ERC), and by U.S. grants ONR-000141110780 and NSF-DMR-1105512. This research was supported in part by the National Science Foundation under Grant No. NSF PHY11-25915.


\begin{thebibliography}{77}
\bibitem{mermin1979} N.D. Mermin, Rev. Mod. Phys. {\bf 51}, 591 (1979).


\bibitem{sondhi1993} S.L. Sondhi, A. Karlhede, S.A. Kivelson, and E.H. Rezayi, \prb {\bf 47}, 16419 (1993).

\bibitem{khawaja2001} U. Al Khawaja and H.T.C. Stoof, Nature (London) {\bf 411}, 918 (2001).

\bibitem{bogdanov1994} A. Bogdanov, A. Hubert, J. Magn. Magn. Mater. \textbf{138}, 255 (1994).

\bibitem{muehlbauer2009} S. M\"uhlbauer, B. Binz, F. Jonietz, C. Pfleiderer, A. Rosch, A. Neubauer, R. Georgii, and P. B\"oni, Science {\bf 323}, 915 (2009).
\bibitem{yu2010} X. Z. Yu, Y. Onose, N. Kanazawa, J. H. Park, J. H. Han, Y. Matsui, N. Nagaosa, and Y. Tokura, Nature {\bf 465}, 09124 (2010)

\bibitem{jonietz2010} F. Jonietz, S. Muehlbauer, C. Pfleiderer, A. Neubauer, W. Muenzer, A. Bauer, T. Adams, R. Georgii, P. Boeni, R. A. Duine, K. Everschor, M. Garst, A. Rosch, Science \textbf{330}, 1648 (2010).

\bibitem{kiselev2011} N. S. Kiselev, A. N. Bogdanov, R. Sch\"afer and U. K. R\"ossler, J. Phys. D {\bf 44}, 392001 (2011).
\bibitem{fert2013} A. Fert, V. Cros, and J. Sampaio, Nature Nanotechnology {\bf 8}, 152 (2013).

\bibitem{neubauer2009} A. Neubauer, C. Pfleiderer, B. Binz, A. Rosch, R. Ritz, P. G. Niklowitz, and P. B\"oni, \prl {\bf 102}, 186602 (2009).

\bibitem{schultz2012} T. Schulz,	 R. Ritz,	 A. Bauer,	 M. Halder,	 M. Wagner,	 C. Franz,	 C. Pfleiderer,	 K. Everschor,	 M. Garst	 and  A. Rosch, Nature Physics {\bf 8}, 301 (2012).

\bibitem{hals2013} Kjetil M. D. Hals and Arne Brataas, \prb {\bf 87}, 174409 (2013).

\bibitem{heinze2011} For another example of skyrmion textures stabilized by interface effects, see S. Heinze, K. von Bergmann, M. Menzel, J. Brede , A. Kubetzka, R. Wiesendanger, G. Bihlmayer and S. Bl\"ugel, Nature Physics {\bf 7}, 2045 (2011).

\bibitem{liu2012} L. Liu, O. J. Lee, T. J. Gudmundsen, D. C. Ralph, R. A. Buhrman,
Phys. Rev. Lett. {\bf 109}, 096602 (2012).

\bibitem{miron2011} I. Miron, K. Garello, G. Gaudin, P.-J. Zermatten, M. Costache, S. Auffret, S. Bandiera, B. Rodmacq, A. Schuhl, and P. Gambardella, Nature (London) {\bf 476}, 189 (2011); I. Miron, T. Moore, H. Szambolics, L. Buda-Prejbeanu, S. Auffret, B. Rodmacq, S. Pizzini, J. Vogel, M. Bonfim, A. Schuhl, and G. Gaudin, Nat. Mater. {\bf 10},  419 (2011).

\bibitem{haazen2012} P. P. J. Haazen,	 E. Mur\'e,	 J. H. Franken,	 R. Lavrijsen,	 H. J. M. Swagten	 and B. Koopmans, Nature Materials {\bf 12}, 299 (2013).
\bibitem{emori2013} S. Emori, U. Bauer, S. Ahn, E. Martinez, G.S.D. Beach, Nature Materials \textbf{12}, 611 (2013).
\bibitem{ryu2013} Kwang-Su Ryu, Luc Thomas, See-Hun Yang and Stuart Parkin,  Nature Nanotechnology \textbf{8}, 527 (2013).

\bibitem{bijl2012} E. van der Bijl and R.A. Duine, \prb {\bf 86}, 094406 (2012).
\bibitem{hals2013b} K.M.D. Hals and Arne Brataas, \prb {\bf 88}, 085423 (2013).
\bibitem{stier2013} Martin Stier, Reinhold Egger, and Michael Thorwart, Phys. Rev. B {\bf 87}, 184415 (2013).


\bibitem{kurebayashi2013} H. Kurebayashi, Jairo Sinova, D. Fang, A. C. Irvine, J. Wunderlich, V. Novak, R. P. Campion, B. L. Gallagher, E. K. Vehstedt, L. P. Zarbo, K. Vyborny, A. J. Ferguson, T. Jungwirth, 	arXiv:1306.1893 [cond-mat.mes-hall].


\bibitem{haney2013} P.M. Haney, Hyun-Woo Lee, Kyung-Jin Lee, Aur\'elien Manchon, and M.D. Stiles, arXiv:13091356.
\bibitem{freimuth2013b} Frank Freimuth, Stefan Blügel, and Yuriy Mokrousov, arXiv:1305.4873 [cond-mat.mtrl-sci].


\bibitem{garate2010} Ion Garate and M. Franz, Phys. Rev. Lett. {\bf 104}, 146802 (2010).
\bibitem{yokoyama2010} Takehito Yokoyama, Jiadong Zang, and Naoto Nagaosa, 	Phys. Rev. B {\bf 81}, 241410(R) (2010).
\bibitem{tserkovnyak2012} Yaroslav Tserkovnyak and Daniel Loss, Phys. Rev. Lett. {\bf 108}, 187201 (2012).

\bibitem{bazaliy1998} Ya. B. Bazaliy, B. A. Jones, and Shou-Cheng
Zhang, \prb {\bf 57}, R3213 (1998).
\bibitem{rossier2004} J. Ferna\'ndez-Rossier, M. Braun, A. S. Nu\'n\~ez, and A. H. MacDonald, \prb {\bf 69}, 174412 (2004).
\bibitem{zhang2004} S. Zhang and Z. Li, Phys. Rev. Lett. {\bf 93}, 127204 (2004).

\bibitem{tserkovnyak2006} Y. Tserkovnyak, H.J. Skadsem, A. Brataas, and G. E. W. Bauer, \prb {\bf 74}, 144405 (2006).
\bibitem{kohno2006} H. Kohno, G. Tatara, and J. Shibata,
J. Phys. Soc. Japan {\bf 75}, 113706 (2006).
\bibitem{piechon2006} F. Pi\'echon and A. Thiaville, \prb {\bf 75}, 174414 (2007).
\bibitem{duine2007} R. A. Duine, A. S. N\'u\~nez, Jairo Sinova, and A. H. MacDonald, Phys. Rev. B {\bf 75}, 214420 (2007).


\bibitem{bijl2013} E. van der Bijl, R.E. Troncoso and R.A. Duine, Phys. Rev. B {\bf 88}, 064417 (2013).


\bibitem{manchon2009} A. Manchon and S. Zhang, Phys. Rev. B {\bf 78}, 212405 (2008); {\bf 79}, 094422 (2009).
\bibitem{garate2009} I. Garate and A. H. MacDonald, Phys. Rev. B {\bf 80}, 134403 (2009).
\bibitem{pesin2012} D. A. Pesin and A. H. MacDonald, Phys. Rev. B {\bf 86}, 014416 (2012).

\bibitem{slonczewski1996} J.C. Slonczewski, J. Magn. Magn. Mater. {\bf 159}, L1 (1996).

\bibitem{makhfudz2012} Imam Makhfudz, Benjamin Krueger, and Oleg Tchernyshyov,  Phys. Rev. Lett. {\bf 109}, 217201 (2012).

\bibitem{sampaio2013} J. Sampaio,	 V. Cros,	 S. Rohart,	 A. Thiaville	 and  A. Fert, Nature Nanotechnology, advance online publication, doi:10.1038/nnano.2013.210 (2013). 


\bibitem{integrals} They are determined by the integrals
\mbox{$D = \frac{1}{4} \int_0^\infty d \rho \rho\left[  \frac{\sin^2 \theta}{\rho^2} + \left(\frac{d\theta}{d\rho}\right)^2\right]$},
\mbox{$I  =   \frac{1}{4} \int_0^\infty d \tilde \rho \left( \sin \theta + \tilde \rho \cos \theta \frac{d\theta}{d\tilde \rho} \right)$},
and \mbox{$ I' = \frac{1}{4} \int_0^\infty d \tilde \rho \left( \sin \theta \cos \theta+ \tilde \rho  \frac{d\theta}{d \tilde \rho} \right)$}.

\bibitem{matrix} Explicitly, its elements are given by $R_{xx} (\phi_0)=\cos \phi_0, R_{xy} (\phi_0)=\sin \phi_0, R_{yx}(\phi_0)=-\sin \phi_0, R_{yy} (\phi_0)=\cos \phi_0$.



\bibitem{integralslattice} In detail, we have that
\mbox{$D_{\alpha\beta} = \int \frac{d\tilde {\bf x}}{4\pi} \frac{\partial \vectomg}{\partial \tilde x_\alpha} \cdot \frac{\partial \vectomg}{\partial\tilde  x_\beta}$},
\mbox{$I_{\alpha\beta}=- \int \frac{d\tilde {\bf x}}{4\pi } \frac{\partial \Omega_\gamma}{\partial \tilde x_\alpha} \epsilon_{\gamma\beta}$},
\mbox{$I'_{\alpha\beta}=- \int \frac{d\tilde {\bf x}}{4\pi } \left(\frac{\partial \vectomg}{\partial \tilde x_\alpha}\times \vectomg \right)_\gamma \epsilon_{\gamma\beta}$}, where all integrals are over the unit cell of the skyrmion lattice, and the lengths are made dimensionless by rescaling $x_\alpha \to \tilde x_\alpha \lambda$.


\bibitem{tserkovnyak2008b} Yaroslav Tserkovnyak and Matthew
Mecklenburg, Phys. Rev. B {\bf 77}, 134407 (2008).

\bibitem{stern1992} Ady Stern, Phys. Rev. Lett. {\bf 68}, 1022 (1992).

\bibitem{barnes2007} S.E. Barnes and S. Maekawa, \prl {\bf 98},
246601 (2007).

\bibitem{duine2008} R.A. Duine, Phys. Rev. B {\bf 77}, 014409 (2008).

\bibitem{tatara2013} Gen Tatara, Noriyuki Nakabayashi, Kyun-Jin Lee, Phys. Rev. B {\bf 87}, 054403 (2013).
\bibitem{yamane2013} Yuta Yamane, Jun'ichi Ieda and Sadamichi Maekawa, \prb {\bf 88}, 014430 (2013). 

\bibitem{tserkovnyak2009} Yaroslav Tserkovnyak and Clement H. Wong, Phys. Rev. B {\bf 79}, 014402 (2009).

\bibitem{zang2011} Jiadong Zang, Maxim Mostovoy, Jung Hoon Han, and Naoto Nagaosa, \prl {\bf 107}, 136804 (2011).

\bibitem{iwasaki2013} Junichi Iwasaki,	 Masahito Mochizuki and Naoto Nagaosa, Nature Nanotech. {\bf 8}, 742 (2013).
    
\bibitem{rohart2013} S. Rohart and A. Thiaville, arXiv:1310.0666 [cond-mat.mes-hall].
\bibitem{romming2013} Niklas Romming, Christian Hanneken, Matthias Menzel, Jessica E. Bickel, Boris Wolter, Kirsten von Bergmann, André Kubetzka, and Roland Wiesendanger, Science {\bf 341}, 636 (2013).

\bibitem{freimuth2013} Frank Freimuth, Robert Bamler, Yuriy Mokrousov, Achim Rosch, 	arXiv:1307.8085 [cond-mat.str-el].

\end{thebibliography}
\end{document}